\documentclass[10pt]{iopart}

\usepackage{iopams}  
\usepackage{graphicx}
\newcommand{\ba}{\begin{eqnarray}}
\newcommand{\ea}{\end{eqnarray}}
\begin{document}
\title{Form factors in the Algebraic Cluster Model}

\author{Roelof Bijker}

\address{Instituto de Ciencias Nucleares, 
Universidad Nacional Aut\'onoma de M\'exico, 
A.P. 70-543, 04510 M\'exico, D.F., M\'exico}
\ead{bijker@nucleares.unam.mx}

\begin{abstract}
I present a derivation of form factors in the Algebraic Cluster Model 
for an arbitrary number of identical clusters. The form factors correspond 
to representation matrix elements which are derived in closed form 
for the harmonic oscillator and deformed oscillator limits. These results 
are relevant for applications in nuclear, molecular and hadronic 
physics.
\end{abstract}

\pacs{03.65.Fd, 21.60.Gx, 13.40.Gp}
%
\vspace{2pc}
\noindent{\it Keywords}: Algebraic cluster model, electromagnetic form factors
%
%
%
\ioptwocol

\section{Introduction}

The dynamics of quantum many-body systems can be studied by applying external probes. 
The response of these systems to strong external fields leads to multiple excitation 
of the target system involving the excitation of many intermediate states. Examples are 
Coulomb excitation \cite{Alder} and medium-energy proton scattering at forward angles 
\cite{Ralph} in nuclear physics, and electron scattering from polar molecules \cite{Collins}. 
The standard approach to treat the coupling between target and projectile to all orders 
is that of a coupled-channel approach, which becomes complicated when the number of 
channels that has to be included is large. An alternative method is based on the 
eikonal or Glauber approximation in which the multiple scattering is summed to all orders 
and which gives a good description of the scattering at forward angles \cite{Ralph}. 

In an eikonal treatment of scattering from complex systems the scattering operator 
is expressed as an exponentiated multipole operator. If the eikonal treatment of the 
scattering problem is combined with an algebraic model of the quantum many-body system, 
the matrix elements of the eikonal scattering operator can be interpreted as representation 
matrix elements which can be obtained 
exactly to all orders in the coupling between target and projectile \cite{Balster,KVI}. 
In addition, there are special solutions in which these matrix elements can be derived 
in closed analytic form \cite{Joe,BAS}.

In nuclear physics, these techniques have been applied to medium-energy proton 
scattering \cite{Joe,KVI1}, Coulomb excitation \cite{KVI2}, subbarrier fusion 
\cite{Baha} and electromagnetic excitation of $\alpha$-cluster nuclei \cite{ACM}, 
and in molecular physics to medium-energy electron scattering from polar molecules 
\cite{BAS}. 

The derivation of representation matrix elements and form factors for systems which are 
dominated by a single multipole, {\it e.g.} quadrupole oscillations in collective nuclei 
and the dipole degree of freedom in polar molecules, was studied for an arbitrary 
multipole in Ref.~\cite{BG}. The aim of this contribution is to study a generalization 
to a system of coupled oscillators as is relevant for $\alpha$-cluster nuclei. 
The results are valid for an arbitrary number of clusters.  

\section{The Algebraic Cluster Model}

The Algebraic Cluster Model (ACM) describes the relative motion of $k$ clusters. 
It is based on the spectrum generating algebra of $U(\nu+1)$ where $\nu=3(k-1)$ 
represents the number of relative spatial degrees of freedom. 
As a special case the ACM contains the $U(4)$ vibron model for two-body problems 
($k=2$) with applications in diatomic molecules \cite{vibron}, nuclear clusters 
\cite{cluster}, and quark-antiquark configurations in mesons \cite{mesons}. 
Further extensions of this idea are the $U(7)$ model for three-body clusters 
($k=3$) with applications to three-quark configurations in baryons \cite{BIL}, 
triatomic molecules \cite{BDL} and nuclear clusters \cite{ACM,C12}, and the 
$U(10)$ model for four-body clusters ($k=4$) which was introduced recently to 
describe the properties of the nucleus $^{16}$O in terms of four-alpha  
clusters \cite{RB,O16}.  

The relevant degrees of freedom of a system of $k$-body clusters are given 
by the $k-1$ relative Jacobi coordinates 
\ba
\vec{\rho}_{j} \;=\; \frac{1}{\sqrt{j(j+1)}} 
\left( \sum_{j'=1}^{j} \vec{r}_{j'} - j \vec{r}_{j+1} \right) ~, 
\label{Jacobi}
\ea
and their conjugate momenta, $\vec{p}_{\rho_{j}}$. Here $\vec{r}_j$ denotes the 
position vector of the $j$-th cluster. The ACM is based on a bosonic quantization 
which consists in introducing $k-1$ vector boson operators (one for each relative 
coordinate) which are related to the coordinates and their conjugate momenta by 
\ba
b^{\dagger}_{j,m} &=& \frac{1}{\sqrt{2}} \left( \rho_{j,m} - ip_{\rho_{j,m}} \right) ~,
\nonumber\\
b_{j,m} &=& \frac{1}{\sqrt{2}} \left( \rho_{j,m} + ip_{\rho_{j,m}} \right) ~,
\ea
with $m=-1,0,1$, and an additional auxiliary scalar boson, $s^{\dagger}$, $s$. 
The set of $(3k-2)^2$ bilinear products of creation and annihilation operators 
generates the Lie algebra of $U(3k-2)$. Since the building blocks of the ACM 
are bosons, all states of the system belong to the totally symmetric representation 
$[N]$ of $U(3k-2)$ where $N$ represents the total number of bosons $N=n_s+\sum_{i} n_{i}$. 

In this contribution, I study the ACM for identical clusters which is relevant to 
$\alpha$-cluster nuclei like $^{12}$C and $^{16}$O. For these systems, the Hamiltonian 
has to be invariant under the permutation group $S_k$ for $k$ identical objects. 
The most general one- and two-body Hamiltonian that describes the relative motion 
of a system of $k$ identical clusters, is a scalar under $S_k$, conserves angular 
momentum and parity as well as the total number of bosons, is given by 
\ba
H &=& \epsilon_{0} \, s^{\dagger} \tilde{s}
- \epsilon_{1} \, \sum_{i} b_{i}^{\dagger} \cdot \tilde{b}_{i} 
+ u_0 \, s^{\dagger} s^{\dagger} \, \tilde{s} \tilde{s}  
\nonumber\\
&& - u_1 \, \sum_{i} s^{\dagger} b_{i}^{\dagger} \cdot \tilde{b}_{i} \tilde{s} 
+ v_0 \, \left( \sum_i b_{i}^{\dagger} \cdot b_{i}^{\dagger} \, \tilde{s} \tilde{s} + {\rm h.c.} \right)
\nonumber\\
&& + \sum_{L} \sum_{iji'j'} v^{(L)}_{iji'j'} \, [ b_{i}^{\dagger} \times b_{j}^{\dagger} ]^{(L)} \cdot 
[ \tilde{b}_{i'} \times \tilde{b}_{j'} ]^{(L)} ~,
\label{HSk}
\ea
with $\tilde{b}_{i,m}=(-1)^{1-m} b_{i,-m}$ and $\tilde{s}=s$. By construction, the  
$\epsilon_0$, $\epsilon_1$, $u_0$, $u_1$ and $v_0$ terms in Eq.~(\ref{HSk}) are invariant under $S_k$. 
The permutation symmetry imposes additional restrictions on the coefficients $v^{(L)}_{iji'j'}$ 
of the last term.  

The energy eigenvalues are obtained numerically by diagonalizing the Hamiltonian in a 
coupled harmonic oscillator basis. The corresponding wave functions are 
characterized by the total number of bosons $N$, angular momentum and parity $L^P$ 
and permutation symmetry $t$. In this contribution it is assumed that the identical clusters 
have no internal structure (like in the application to $\alpha$-cluster nuclei). As a consequence, 
the wave functions have to be completely symmetric under the permutation group $S_k$  

The Algebraic Cluster Model has a rich algebraic structure, which includes 
both continuous and discrete symmetries. It is of general interest to study 
limiting cases of the Hamiltonian of Eq.~(\ref{HSk}), in which the energy 
spectra and form factors can be obtained in closed form. In this contribution I consider 
two dynamical symmetries of the ACM Hamiltonian for the $k$-body problem 
which are related to the group lattice
\ba
U(3k-2) \supset \left\{ \begin{array}{c} U(3k-3) \\ \\ SO(3k-2) \end{array} \right\} 
\supset SO(3k-3) ~,
\ea
which are called the $U(3k-3)$ and $SO(3k-2)$ limits of the ACM, respectively.  
A geometric analysis shows that the $U(3k-3)$ limit corresponds for large $N$ 
to the (an)harmonic oscillator in $3(k-1)$ dimensions and the $SO(3k-2)$ limit 
to the deformed oscillator in $3(k-1)$ dimensions \cite{RB}. 

\section{Transition form factors}

Transition probabilities, charge radii, and other electromagnetic properties of 
interest can be obtained from the transition form factors. For electric transitions 
the form factors correspond to the matrix elements of the Fourier transform of the 
charge distribution
\ba
F(\vec{q}) &=& \int d\vec{r} \, 
e^{i\vec{q} \cdot \vec{r}} \, \langle \alpha' L' M' \, 
| \, \hat{\rho}(\vec{r}) \, | \, \alpha L M \rangle ~.
\ea
For an extended charge distribution in which the charges of the clusters 
are smeared by a Gaussian
\ba
\rho (\vec{r}) &=& \frac{Ze}{k} \left( \frac{\gamma}{\pi} \right)^{3/2} 
\sum_{i=1}^{k} e^{-\gamma \left( \vec{r}-\vec{r}_{i} \right)^{2}} ~, 
\ea
the transition form factor reduces to 
\ba
F(\vec{q}) &=& Ze \sum_{M''} {\cal D}^{(L')}_{M'M''}(\hat{q}) 
\, {\cal F}(q) \, {\cal D}^{(L)}_{M''M}(-\hat{q})  ~, 
\ea
with
\ba
{\cal F}(q) \;=\; e^{-q^2/4\gamma} \langle \alpha' L' M'' \, 
| \, e^{i \vec{q} \cdot \vec{r}_{k}} \, | \, \alpha L M'' \rangle 
\nonumber\\
\hspace{0.5cm} \;=\; e^{-q^2/4\gamma} \langle \alpha' L' M'' \, 
| \, e^{-i q \sqrt{\frac{k-1}{k}} \rho_{k-1,z}} \, | \, \alpha L M'' \rangle ~.
\label{calff}
\ea 
In the derivation I have used the symmetry of the wave functions, made a 
transformation to Jacobi coordinates and integrated over the center-of-mass 
coordinate.  

In the ACM, these matrix elements can be obtained algebraically by making the replacement 
\ba
\sqrt{(k-1)/k} \, \rho_{k-1,z} \;\rightarrow\; q \beta D_{k-1,0}/X_{D} ~, 
\ea
where
\ba
D_{k-1,0} \;=\; b_{k-1,0}^{\dagger} \, s + s^{\dagger} \, b_{k-1,0} ~.
\ea
The coefficient $X_{D}$ is a normalization factor and is equal to the reduced matrix element 
of the dipole operator between the ground state with $L^P=0^+$ and the first excited state 
with $L^P=1^-$  
\ba
X_D \;=\; \left< 1^-_1 || D_{k-1} || 0^+_1 \right> ~.
\ea
Therefore, in the ACM one has to evaluate the matrix elements of 
the transition operator 
\ba
\hat U(\epsilon) &=& e^{i \epsilon \hat D_{k-1,0}} ~,
\ea
with $\epsilon=-q \beta/X_D$ which can be interpreted as representation 
matrix elements of $U(3k-2)$, {\it i.e.} generalizations of the Wigner 
$D$-functions for $SU(2)$. In general, these matrix elements can be derived 
from the transformation properties of a single boson. The transition operator 
$U(\epsilon)$ transforms the scalar boson and the $z$-component of the 
($k-1$)-th Jacobi boson amongst each other, and does not affect the other bosons.  
\ba
U(\epsilon) \left( \begin{array}{c} s^{\dagger} \\ b_{k-1,0}^{\dagger} \end{array} \right) U(-\epsilon) 
\nonumber\\
\hspace{1cm} \;=\; \left( \begin{array}{cc} \cos \epsilon & i \sin \epsilon \\
i \sin \epsilon & \cos \epsilon \end{array} \right) 
\left( \begin{array}{c} s^{\dagger} \\ b_{k-1,0}^{\dagger} \end{array} \right) ~.
\label{boson}
\ea
There are special solutions of the ACM in which these matrix elements can be derived in closed form. 
These solutions correspond to dynamical symmetries of the ACM Hamiltonian. Here I discuss two of them: 
the $U(3k-3)$ limit (harmonic oscillator) and the $SO(3k-2)$ limit (deformed oscillator). 
The general procedure to obtain the form factors for a single $(2\lambda+1)$-dimensional oscillator 
was outlined in Ref.~\cite{BG}, and will be generalized here to the ACM for a system of $k-1$ coupled 
three-dimensional oscillators. 

\subsection{Harmonic oscillator}

In the absence of the $v_0$ term in Eq.~(\ref{HSk}), there is no coupling between different 
harmonic oscillator shells. The oscillator is harmonic if all terms, except $\epsilon_0$ and 
$\epsilon_1$, are set to zero; otherwise it is anharmonic. This dynamical 
symmetry corresponds to the group reduction 
\ba
\left| \begin{array}{ccccc}
U(3k-2) &\supset& U(3k-3) &\supset& SO(3k-3) \\
\, [N] &,& n &,& \tau \end{array} \right. 
\nonumber\\
\hspace{1.5cm} \left. \begin{array}{ccccc}
&\supset& SO(3) &\supset& SO(2) \\ 
&,\alpha,& L_t &,& M \end{array} \right> 
\ea
The label $n$ represents the total number of oscillator quanta $n=\sum_i n_i=0,1,\ldots,N$. 
The energy levels are grouped into oscillator shells characterized by $n$ and parity $P=(-1)^{n}$. 
The levels belonging to an oscillator shell are further classified by the symmetric irreducible 
representation $\tau$ of $SO(3k-3)$ with $\tau=n,n-2,\ldots,1$ or $0$ for $n$ odd or even, 
the angular momentum $L$ and its projection $M$, and the permutation symmetry $t$. 
$\alpha$ denotes all additional labels that are needed for a unique classification scheme. 
This special case is called the $U(3k-3)$ limit of the ACM. 

The wave functions for the $U(3k-3)$ limit are given by
\ba
\left| [N]n \tau \alpha L_t M \right> &=& \frac{B_{n\tau}(s^{\dagger})^{N-n}}{\sqrt{(N-n)!}} \, 
\left( \sum_{i=1}^{k-1} b^{\dagger}_{i} \cdot b^{\dagger}_{i} \right)^{\frac{n-\tau}{2}} 
\nonumber\\ 
&& \times \left| [\tau]\tau \tau \alpha L_t M \right> ~,
\ea
with
\ba
B_{n\tau} \;=\; (-)^{\frac{n-\tau}{2}} \, \sqrt{ \frac{(2\tau+3k-5)!!}{(n+\tau+3k-5)!!(n-\tau)!!} } ~.
\ea

The matrix elements of the transition operator $\hat U(\epsilon)$ can be derived by using the 
transformation properties of Eq.~(\ref{boson})
\ba 
U_{Nn \tau \alpha L_t}(\epsilon) \;=\; \left< [N]n \tau \alpha L_t M=0 \right| \hat U(\epsilon) \left| [N]00000 \right> 
\nonumber\\
\hspace{0.5cm} \;=\; B_{n\tau} \, A_{\tau \alpha L_t} \, \sqrt{\frac{N!}{(N-n)!}} \, 
(i \sin \epsilon)^n \, (\cos \epsilon)^{N-n} 
\ea
with 
\ba
A_{\tau \alpha L_t} &=& \frac{1}{\tau!} \, 
\left< [\tau] \tau \tau \alpha L_t 0 \right| ( b_{k-1,0}^{\dagger} )^{\tau}  \left| [0]00000 \right> ~.
\ea
In general, the coefficients $A_{\tau \alpha L_t}$ have to calculated explicitly. Only for the case of two-body 
clusters they have been derived in closed form \cite{BAS}. 

For large $N$, the $U(3k-3)$ limit corresponds to the (an)harmonic oscillator \cite{RB}. 
The coefficient $\epsilon$ is given by $\epsilon=-q \beta/X_D$ with $X_D=\sqrt{3N}$. In the  
large $N$ limit which is taken such that $n/N \ll 1$ and $q\beta$ remains finite,  
the transition matrix elements reduce to
\ba 
U_{Nn \tau \alpha L_t}(\epsilon) 
&\rightarrow& B_{n\tau} \, A_{\tau \alpha L_t} \, 
\left( \frac{-iq\beta}{\sqrt{3}} \right)^n \, e^{-q^2\beta^2/6} ~.
\ea

\begin{figure}
\includegraphics[scale=0.33]{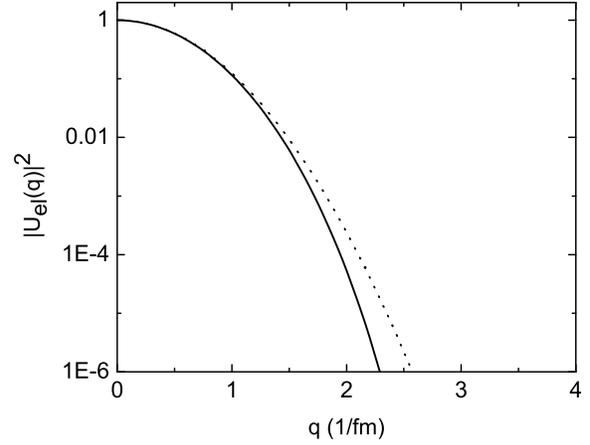}
\caption[]
{Elastic form factor in the $U(3k-3)$ limit of the ACM calculated with   
$\beta=2.50$ fm for $N=10$ (solid line) and in the large $N$ limit 
(dotted line).}
\label{hosc}
\end{figure}

As an example, the elastic form factor is given by 
\ba
U_{\rm el}(\epsilon) &=& (\cos \epsilon)^{N} 
\nonumber\\ 
&\rightarrow& e^{-q^2\beta^2/6} \;=\; 1 - \frac{1}{6} q^2 \beta^2 + \ldots 
\ea
Fig.~\ref{hosc} shows a comparison of the elastic form factor in the $U(3k-3)$ limit 
calculated for finite $N(=10)$ and in the large $N$ limit.  
The scale parameter $\beta$ is related to the rms radius of the system  
\ba
\left< r^2 \right> \;=\; -6 \left. \frac{{\cal F}(0^+ \rightarrow 0^+;q)}{dq^2} \right|_{q=0} 
\;=\; \beta^2 + \frac{3}{2\gamma} ~. 
\label{rms}
\ea

The probability that a state belonging to a given oscillator shell $n$ 
can be excited from the ground state with $n=0$ is given by the binomial distribution
\ba
P_{n}(\epsilon) &=& \sum_{\tau \alpha L_t} | U_{Nn \tau \alpha L_t}(\epsilon) |^2 
\nonumber\\
&=& \left( \begin{array}{c} N \\ n \end{array} \right) (\sin^2 \epsilon)^n (\cos^2 \epsilon)^{N-n} ~.
\ea
For $\epsilon=0$ only the ground state is excited. With increasing 
values of $\epsilon$ all higher oscillator shells are successively excited until for $\epsilon=\pi/2$ 
all strength is concentrated in the highest oscillor shell with $n=N$. The excitation probability 
is symmetric around $\epsilon=\pi/2$, and is a periodic function with period $\pi$ which implies 
that for $\epsilon=\pi$ all strength is again concentrated in the ground state. This behavior is 
an artefact of the finiteness of the model space. The range of $q$ values shown in the figures is up 
to $q=4$ (1/fm) which corresponds to $\epsilon=-q\beta/\sqrt{3N}=-1.83$. The excitation probabilities 
of all states of a given oscillator shell show the same dependence on $\epsilon$ (or $q$), the only 
difference is in the numerical factor $B_{n\tau} \, A_{\tau \alpha L_t}$. 

\begin{figure}
\vfill 
\begin{minipage}{.5\linewidth}
\includegraphics[scale=0.3]{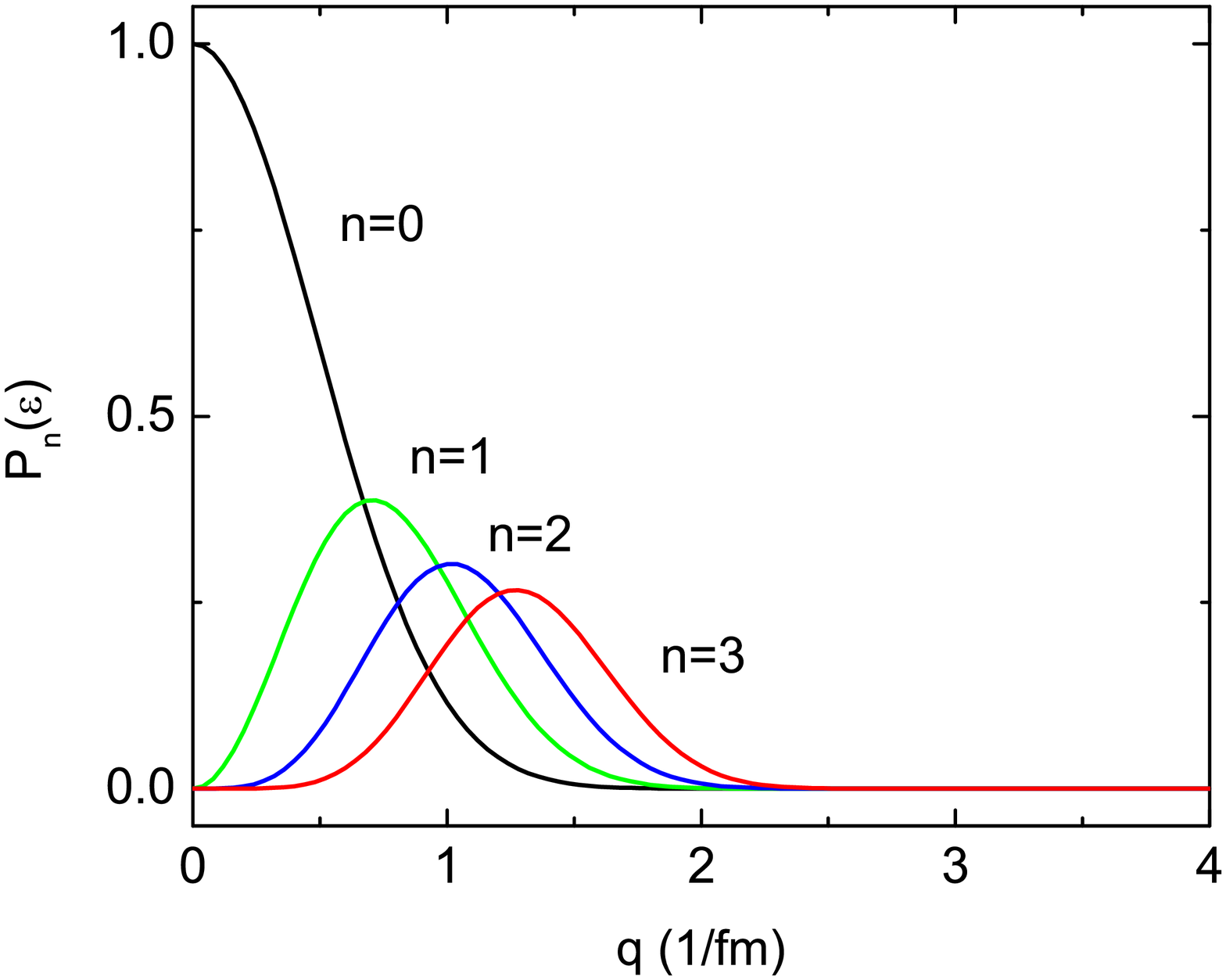}
\end{minipage}\vfill
\begin{minipage}{.5\linewidth}
\includegraphics[scale=0.3]{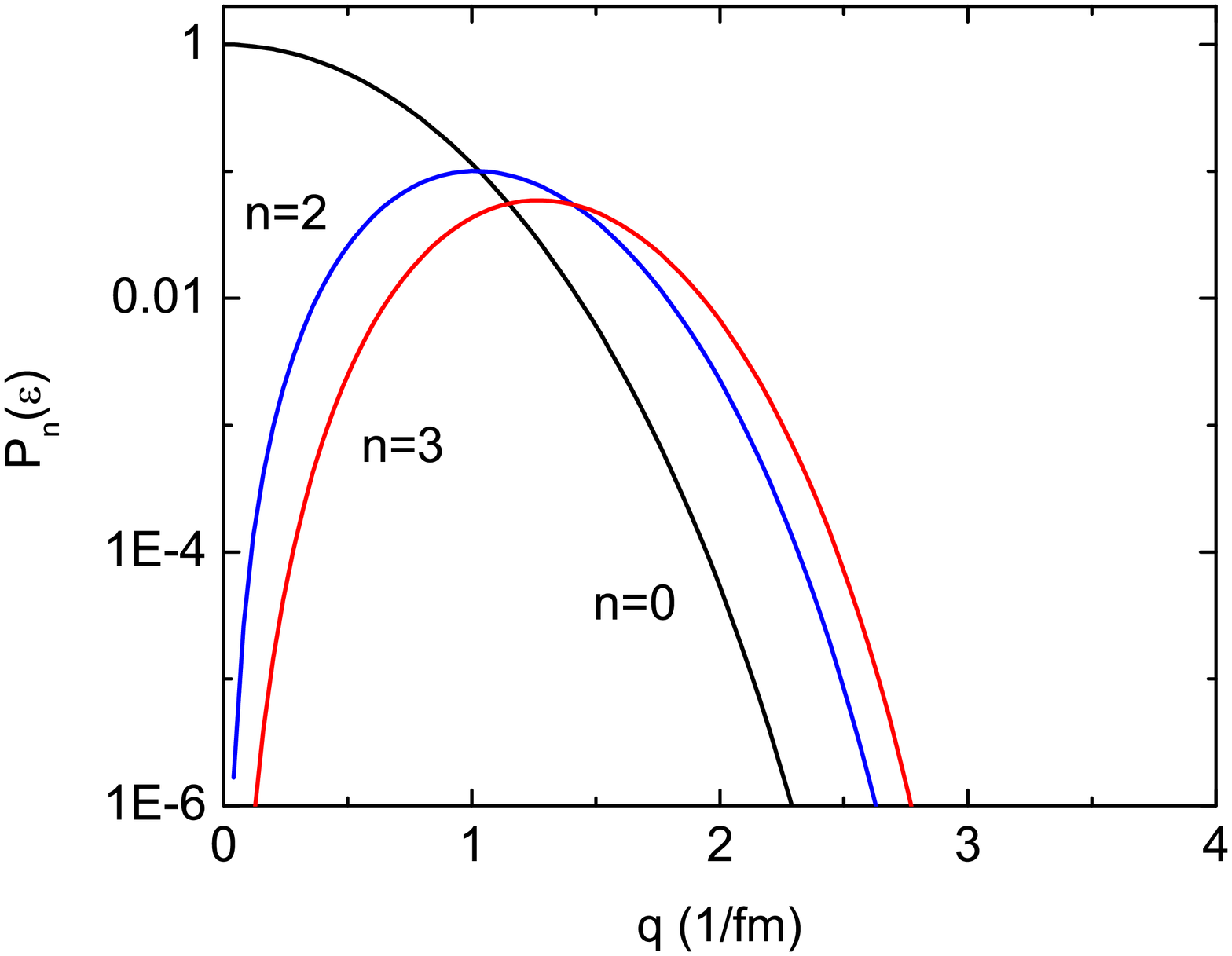}
\end{minipage}
\caption[]
{Excitation probability in the $U(3k-3)$ limit of the ACM (harmonic oscillator) 
for $k=4$ clusters calculated with $\beta=2.50$ fm and $N=10$ for $n=0$ (solid black line), 
$n=1$ (solid green line), $n=2$ (solid blue line) and $n=3$ (solid red line). 
In the top panel the sum is performed over all states of a given oscillator shell, 
whereas in the bottom panel only the symmetric states are taken into account.}
\label{hosc023}
\end{figure}

In the large $N$ limit, the excitation probability reduces to the familiar Poisson distribution 
for the harmonic oscillator \cite{Alder} 
\ba
P_{n}(\epsilon) &\rightarrow& \frac{1}{n!} \left( \frac{q^2\beta^2}{3} \right)^n \, e^{-q^2\beta^2/3} ~. 
\label{Poisson}
\ea
Fig.~\ref{hosc023} shows the results for $P_n$ for the case of four-cluster systems. 
The top panel shows the result for the sum over all states according to Eq.~(\ref{Poisson}). 
In the bottom panel, the sum is restricted to states which are symmetric ($t=[k]$) under 
the permutation group $S_k$, as is relevant for the case of $\alpha$-cluster nuclei. The 
curves for a given oscillator shell have the same shape as in the top panel, but they 
are multiplied by a factor of $1$ for $n=0$, $1/3$ for $n=2$ and $2/9$ for $n=3$. 
The $n=1$ shell is absent since it does not contain a symmetric state. 

\subsection{Deformed oscillator}

For the (an)harmonic oscillator, the number of oscillator quanta $n$ is a good 
quantum number. However, when $v_{0}\neq 0$ in Eq.~(\ref{HSk}), 
the oscillator shells with $\Delta n=\pm 2$ are mixed, and the
eigenfunctions are spread over many different oscillator shells. A dynamical 
symmetry that involves the mixing between oscillator shells, is provided by
the reduction 
\ba
\left| \begin{array}{ccccc}
U(3k-2) &\supset& SO(3k-2) &\supset& SO(3k-3) \\
\, [N] &,& \sigma &,& \tau \end{array} \right.
\nonumber\\
\hspace{1.5cm} \left. \begin{array}{ccccc} 
&\supset& SO(3) &\supset& SO(2) \\
&,\alpha,& L_t &,& M \end{array} \right> ~.
\ea
The label $\sigma =N,N-2,\ldots,1$ or $0$ for $N$ odd or even, respectively, 
characterizes the symmetric representations of $SO(3k-2)$, and 
$\tau=0,1,\ldots,\sigma$ those of $SO(3k-3)$.  The remaining quantum numbers 
are the same as for the harmonic oscillator. 

The wave functions for the $SO(3k-2)$ limit are given by
\ba
\left| [N] \sigma \tau \alpha L_t M \right> = B_{N\sigma} \, ( P^{\dagger} )^{\frac{N-\sigma}{2}} \, 
\left| [\sigma] \sigma \tau \alpha L_t M \right> ~,
\ea
with
\ba
B_{N\sigma}= (-)^{\frac{N-\sigma}{2}} \, \sqrt{ \frac{(2\sigma+3k-4)!!}{(N+\sigma+3k-4)!!(N-\sigma)!!} } ~.
\ea
and $P^{\dagger}$ is the pair creation operator in the boson space
\ba
P^{\dagger} = s^{\dagger}s^{\dagger} - \sum_{i=1}^{k-1} b^{\dagger}_{i} \cdot b^{\dagger}_{i} ~.
\ea
The state with $N=\sigma$ can be written as
\ba
\left| [\sigma] \sigma \tau \alpha L_t M \right> &=& \sum_{j=0}^{[(\sigma-\tau)/2]}
F_{j}(\sigma,\tau) \, ( s^{\dagger} )^{\sigma-\tau-2j} 
\nonumber\\ 
&& \hspace{1.5cm} ( P^{\dagger} )^{j} \, \left| [\tau] \tau \tau \alpha L_t M \right> ~,
\ea
with
\ba
F_{j}(\sigma,\tau) &=& \sqrt{ \frac{(\sigma-\tau)!(2\tau+3k-5)!!}
{(2\sigma+3k-6)!!(\sigma+\tau+3k-5)!}} 
\nonumber\\
&& \hspace{1cm} \left( -\frac{1}{2} \right)^{j} \, \frac{(2\sigma+3k-6-2j)!!}{(\sigma-\tau-2j)!j!} ~.
\ea

For the calculation of transition form factors one has to derive the matrix 
elements of $\hat U(\epsilon)$ between the ground state and an arbitrary 
final state. Since the $\hat D_{k-1}$ is a generator of $SO(3k-2)$ the
transition operator only connects to states which belong to the 
$SO(3k-2)$ ground state band with $\sigma=N$. 
The matrix element for the excitation from the ground state
can be expressed in terms of a Gegenbauer polynomial \cite{BG}
\ba
U_{NN \tau \alpha L_t}(\epsilon) \;=\; 
\left< [N] N \tau \alpha L_t M=0 \left| \hat U(\epsilon) \right| [N] 00000 \right> 
\nonumber\\
\hspace{0.5cm} \;=\; A_{\tau \alpha L_t} \, \sqrt{ \frac{(2\tau+3k-6)!!}{(3k-6)!!} }
\nonumber\\
\hspace{1cm} \sqrt{ \frac{N!(3k-5)!}{(N+3k-5)!} 
\frac{(N-\tau)!(2\tau+3k-5)!}{(N+\tau+3k-5)!} } \,  
\nonumber\\
\hspace{1cm} (i \sin \epsilon)^{\tau} \, 
C_{N-\tau}^{(\tau+\frac{3k-4}{2})}(\cos \epsilon) ~. 
\ea

In the large $N$ limit, the $SO(3k-2)$ limit corresponds to a deformed oscillator 
in $3(k-1)$ dimensions \cite{RB}. 
The coefficient $\epsilon$ is given by $\epsilon=-q \beta/X_D$ with $X_D=\sqrt{N(N+3k-4)/(k-1)}$. 
In the  large $N$ limit which is taken such that $\tau/N \ll 1$ and $q\beta$ remains finite, 
the transition matrix element can be expressed in terms of a spherical Bessel function for 
$k$-body clusters with $k$ even
\ba
U_{NN \tau \alpha L_t}(\epsilon) &\rightarrow& 
A_{\tau \alpha L_t} \, \sqrt{ (3k-5)!!(2\tau+3k-5)!! } 
\nonumber\\
&& \frac{(-i)^{\tau} j_{\tau+\frac{3k-6}{2}}(q\beta \sqrt{k-1})}
{(q\beta \sqrt{k-1})^{\frac{3k-6}{2}}} ~,
\ea
and in terms of a cylindrical Bessel function for $k$ odd
\ba
U_{NN \tau \alpha L_t}(\epsilon) &\rightarrow& 
A_{\tau \alpha L_t} \, \sqrt{ (3k-5)!!(2\tau+3k-5)!! } 
\nonumber\\ 
&&  \frac{(-i)^{\tau} J_{\tau+\frac{3k-5}{2}}(q\beta \sqrt{k-1})}
{(q\beta \sqrt{k-1})^{\frac{3k-5}{2}}} ~.
\ea

\begin{figure}
\vfill 
\begin{minipage}{.5\linewidth}
\includegraphics[scale=0.3]{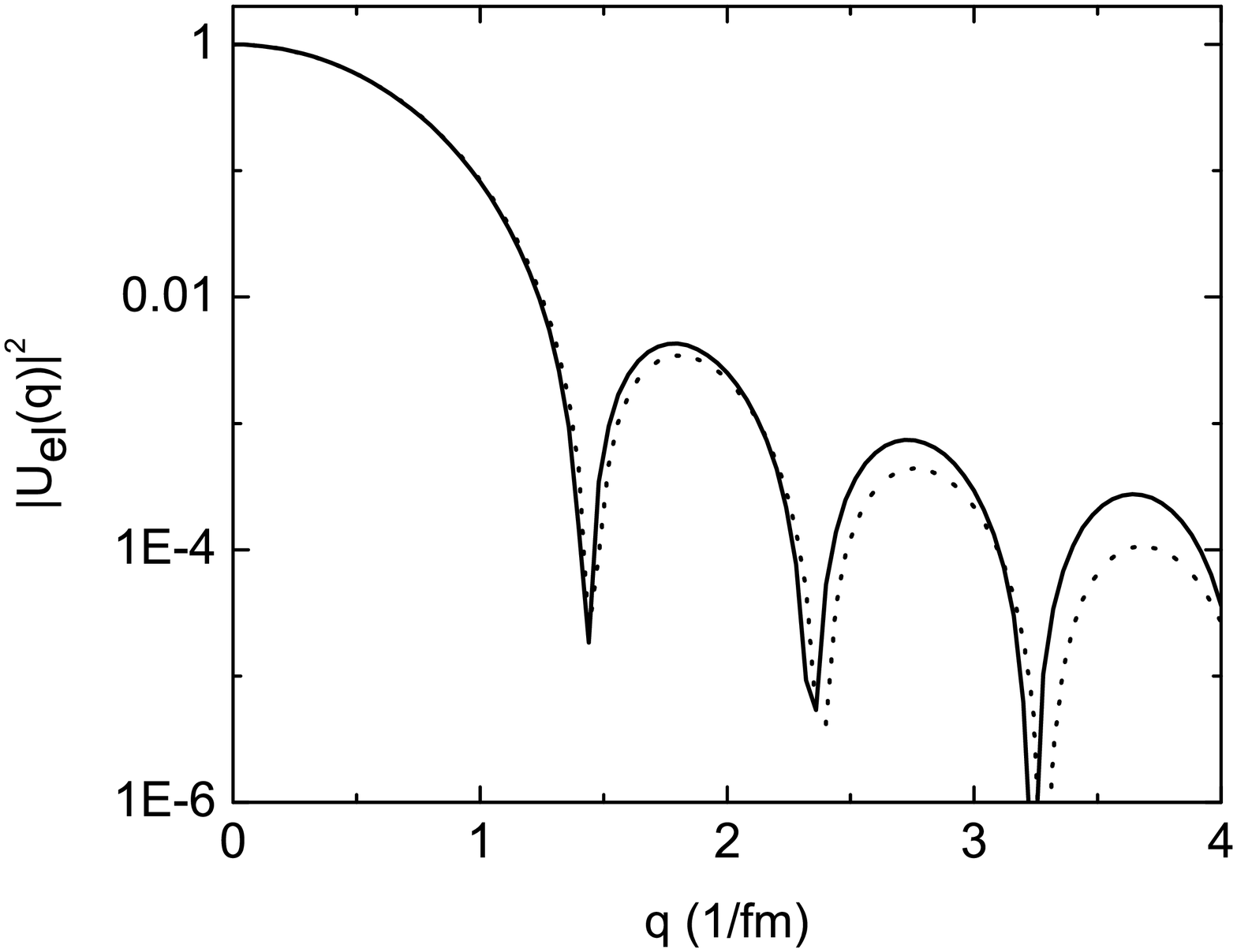}
\end{minipage}\vfill
\begin{minipage}{.5\linewidth}
\includegraphics[scale=0.3]{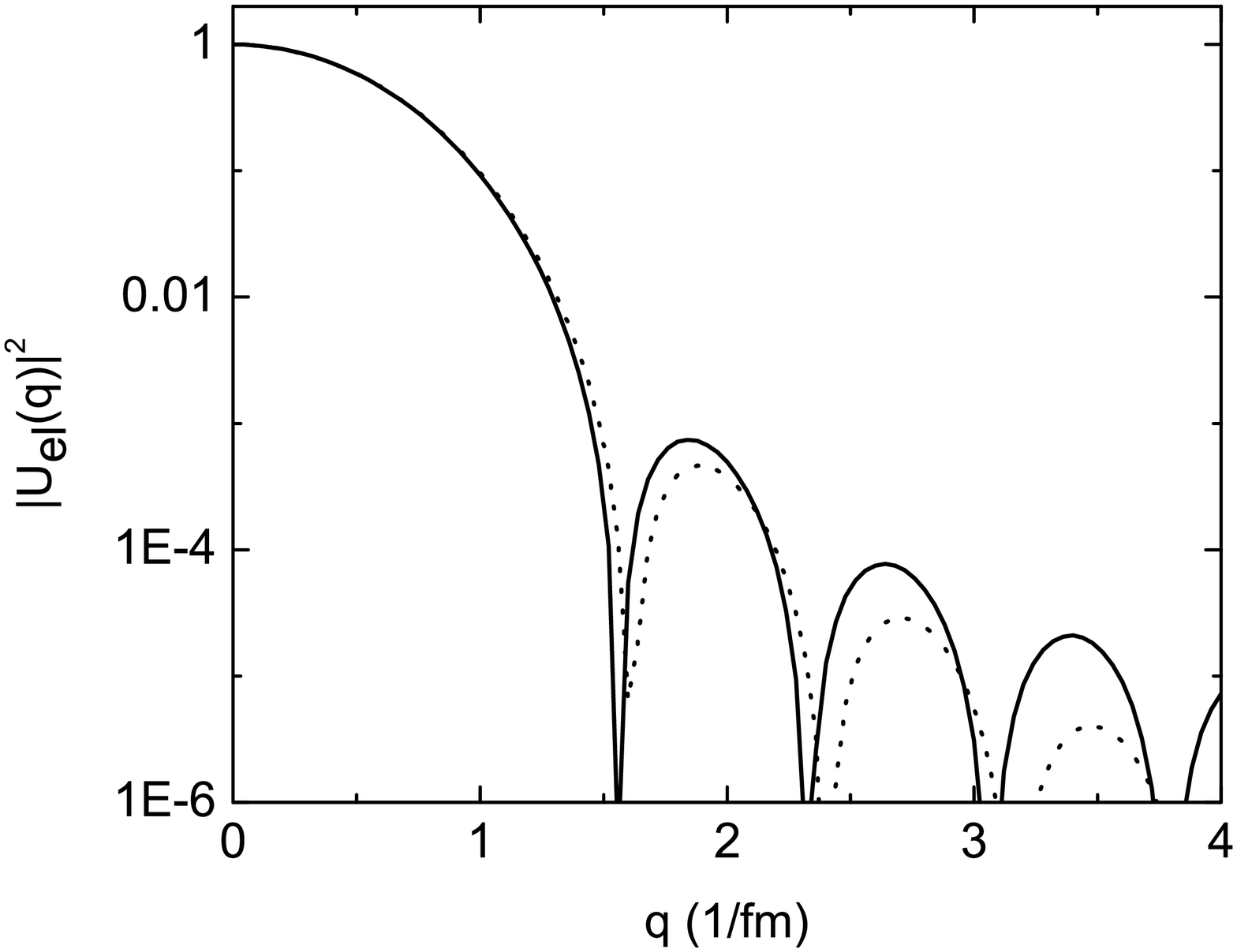}
\end{minipage}
\caption[]
{Elastic form factor in the $SO(3k-2)$ limit of the ACM  
calculated with $\beta=2.50$ fm for $N=10$ (solid line) and in the large $N$ limit 
(dotted line) for three-body clusters (top) and four-body clusters (bottom).}
\label{dosc}
\end{figure}

In the $SO(3k-2)$ limit, the elastic form factor is given by 
\ba
U_{\rm el}(\epsilon) &=& \frac{N!(3k-5)!}{(N+3k-5)!} \, C_{N}^{\frac{3k-4}{2}}(\cos \epsilon)
\nonumber\\
&\rightarrow& \left\{ \begin{array}{cc} 
(3k-5)!! \frac{j_{\frac{3k-6}{2}}(q\beta \sqrt{k-1})}{(q\beta \sqrt{k-1})^{\frac{3k-6}{2}}} 
& \mbox{  for } k \mbox{ even} \\ \\
(3k-5)!! \frac{J_{\frac{3k-5}{2}}(q\beta \sqrt{k-1})}{(q\beta \sqrt{k-1})^{\frac{3k-5}{2}}} 
& \mbox{  for } k \mbox{ odd} 
\end{array} \right. 
\label{uel}
\ea
Fig.~\ref{dosc} shows a comparison of the elastic form factor 
in the $SO(3k-2)$ limit calculated for finite $N$ and in the large $N$ limit for 
three- and four-body clusters, respectively. In this case, the elastic form factor 
shows an oscillatory behavior. The results in the large $N$ limit are 
closer to the exact calculations for the case of three-body clusters than they are 
for four-body clusters. This behavior can be understood qualitatively by first 
expressing the Gegenbauer polynomial in terms of a hypergeometric function and 
next making an expansion in powers of $q \beta$  
\ba
U_{\rm el}(\epsilon) 
&=& _{2}F_{1}(-\frac{N}{2},\frac{N+3k-4}{2},\frac{3k-3}{2};\sin^2 \epsilon)
\nonumber\\
&=& 1 - \frac{1}{6} (q \beta)^2 + \frac{k-1}{24(3k-1)} (q \beta)^4 
\nonumber\\
&& \times \left( 1-\frac{2(3k-4)}{3N^2} + {\cal O}(\frac{1}{N^3}) \right)  - \ldots
\ea
In the large $N$ limit, one has
\ba
U_{\rm el}(\epsilon) 
&\rightarrow& \sum_{n=0}^{\infty} (-1)^n \frac{(k-1)^n (3k-5)!!}{(2n)!!(2n+3k-5)!!} (q \beta)^{2n} 
\nonumber\\
&=& 1 - \frac{1}{6} (q \beta)^2 + \frac{k-1}{24(3k-1)} (q \beta)^4 - \ldots
\ea 
Whereas up to first order in $(q \beta)^2$ the results do not depend on $N$, for the second 
order term there is a $N$ dependent factor that moreover depends on the number of clusters $k$, 
which is smaller for three than for four clusters. The above equations show that also in this case, 
the scale parameter $\beta$ is related to the rms radius according to Eq.~(\ref{rms}). 

Since the dipole operator is a generator of $SO(3k-2)$ only states belonging to 
the ground state band with $\sigma=N$ can be excited from the ground state. 
The probability that a state belonging to the $\tau$ multiplet of the ground state 
band can be excited from the ground state with $\tau=0$ is given by
\ba
P_{\tau}(\epsilon) \;=\; \sum_{\alpha L_t} | U_{NN \tau \alpha L_t}(\epsilon) |^2 ~.
\ea
Just as for the harmonic oscillator, with increasing values of $\epsilon$ all 
higher $\tau$ multiplets are successively excited. The symmetry properties of $P_{\tau}$ are 
the same as for the harmonic oscillator. The range shown in the figures corresponds to 
$\epsilon=-q\beta/\sqrt{N(N+5)/2}=-1.15$ for three-body clusters 
and $\epsilon=-q\beta/\sqrt{N(N+8)/3}=-1.29$ for four-body clusters. 
The excitation probabilities of all states of a given $\tau$-multiplet show the same dependence 
on $\epsilon$ (or $q$), the only difference is in the numerical factor $A_{\tau \alpha L_t}$. 

\begin{figure}
\vfill 
\begin{minipage}{.5\linewidth}
\includegraphics[scale=0.3]{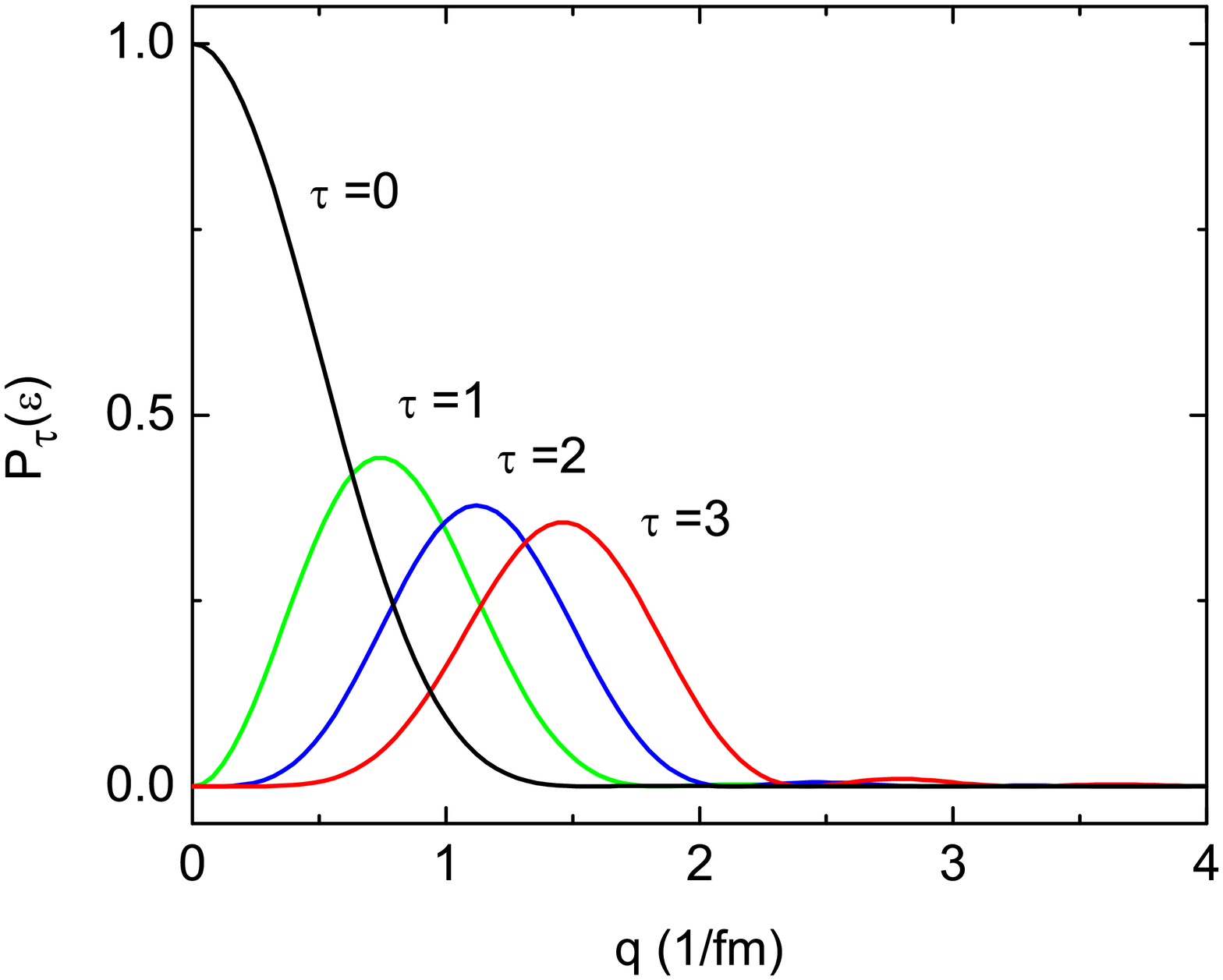}
\end{minipage}\vfill
\begin{minipage}{.5\linewidth}
\includegraphics[scale=0.3]{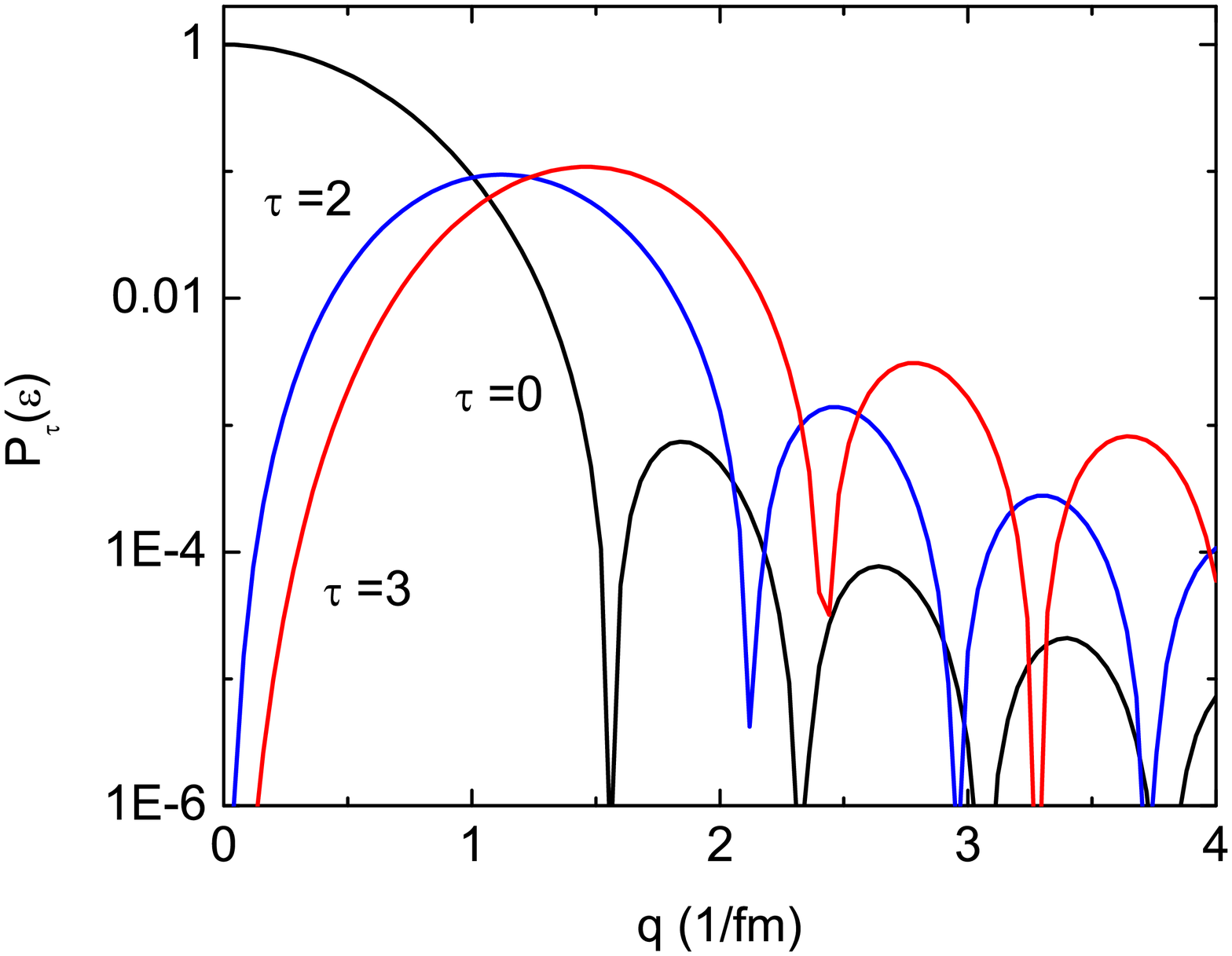}
\end{minipage}
\caption[]
{Excitation probability in the $SO(3k-2)$ limit of the ACM (deformed oscillator) 
for $k=4$ clusters calculated with $\beta=2.50$ fm and $N=10$ for $\tau=0$ (solid black line), 
$\tau=1$ (solid green line), $\tau=2$ (solid blue line) and $\tau=3$ (solid red line). 
In the top panel the sum is performed over all states of a given $\tau$-multiplet, 
whereas in the bottom panel only the symmetric states are taken into account.}
\label{dosc023}
\end{figure}

In the large $N$ limit, the excitation probability reduces to 
\ba
P_{\tau}(\epsilon) &\rightarrow& \frac{(3k-5)!!(2\tau+3k-5)(\tau+3k-6)!}{(3k-6)!!\tau!} 
\nonumber\\
&& \left( \frac{j_{\tau+\frac{3k-6}{2}}(q\beta \sqrt{k-1})}{(q\beta \sqrt{k-1})^{\frac{3k-6}{2}}} \right)^2 ~,
\ea
for $k$ even and
\ba
P_{\tau}(\epsilon) &\rightarrow& \frac{(3k-5)!!(2\tau+3k-5)(\tau+3k-6)!}{(3k-6)!!\tau!} 
\nonumber\\
&& \left( \frac{J_{\tau+\frac{3k-5}{2}}(q\beta \sqrt{k-1})}{(q\beta \sqrt{k-1})^{\frac{3k-5}{2}}} \right)^2 ~,
\ea
for $k$ odd. 
Fig.~\ref{dosc023} shows the results for $P_{\tau}$ for the case of four-cluster systems. 
The top panel shows the result for the sum over all states, whereas in the bottom panel the sum 
is restricted to states which are symmetric ($t=[k]$) under the permutation group $S_k$. 
The curves for a given $\tau$-multiplet have the same shape as in the top panel, but they 
are multiplied by a factor of $1$ for $\tau=0$, $1/4$ for $\tau=2$ and $11/36$ for $\tau=3$. 
The $\tau=1$ multiplet is missing since it contains no symmetric states. 

\section{Summary and conclusions}

In this contribution, I showed how the derivation of transition form factors for systems 
of two- and three-body clusters can be generalized to an arbitrary number of $k$ clusters. 
The derivation was carried out in explicit form for two dynamical symmetries of the 
Algebraic Cluster Model both for finite systems (finite number of bosons $N$) and 
infinite systems (large $N$ limit). The ACM is based on the algebraic quantization 
of the relative Jacobi variables for few-body systems. The ensuing $U(3k-2)$ spectrum 
generating algebra incorporates all vibrational and rotational degrees of freedom from 
the beginning, and takes into account the permutation symmetry of identical clusters in 
an exact manner. 

First I discussed the $U(3k-2) \supset U(3k-3)$ limit which corresponds to the harmonic 
oscillator. With increasing value of the coupling strength $\epsilon$ the different 
oscillator shells are excited successively before they fall off exponentially 
(in the large $N$ limit). The relative transition matrix elements to states belonging 
to the same oscillator shell only depend on a geometric factor ($B_{n\tau} A_{\tau \alpha L}$) 
and not on the coupling strength $\epsilon$. Similarly, in the $SO(3k-2)$ limit 
(deformed oscillator) the different $\tau$ multiplets are excited successively. 
In this case the form factors show an oscillatory behavior since in the large $N$ limit 
they are given by Bessel functions. Just as for the harmonic oscillator, the relative 
matrix elements to states belonging to the same $\tau$ multiplet only depend on a 
geometric factor ($A_{\tau \alpha L}$). 

The present results for transition form factors are of general interest since the ACM has 
found interesting applications in many different areas of physics. Future work includes 
possible applications of the ACM for four-body systems in molecular physics 
(X$_4$ molecules), nuclear physics ($^{16}$O as a cluster of four $\alpha$ particles). 
and hadronic physics ($q^4-\bar{q}$ multiquark configurations). 
As a final comment, it is important to stress that the ACM provides a general framework 
to study the full rotational and vibrational structure of many-body systems which is not 
restricted to the case of identical particles discussed in this contribution. It can 
be applied to other situations as well, such as nonidentical particles 
and/or other geometric configurations \cite{BL}. 

\ack
It is a pleasure and an honor to dedicate this contribution to the careers of 
Margarita and Vladimir Man'ko. We have met on numerous occasions both at UNAM and at the 
{\it Symmetries in Science} meetings in Bregenz, brought together by a shared interest 
in symmetries, group theory and its applications in science. Many congratulations!

This work was supported in part by grant IN107314 from PAPIIT-UNAM.

\end{document}